\documentclass{article} \usepackage{epsfig,latexsym,graphicx}
\usepackage{amsmath}
\usepackage{url}
\usepackage{slashed}
\def\circa#1{\,\raise.3ex\hbox{$#1$\kern-.75em\lower1ex\hbox{$\sim$}}\,}

   \newcommand{\be}{\begin{equation}}
\newcommand{\ee}{\end{equation}} \newcommand{\ben}{\begin{displaymath}}
\newcommand{\een}{\end{displaymath}} \newcommand{\ba}{\begin{eqnarray}}
\newcommand{\ea}{\end{eqnarray}} \newcommand{\ban}{\begin{eqnarray*}}
\newcommand{\ean}{\end{eqnarray*}} 
 
 \newcommand{\pvet}{\mbox{\boldmath $p$}}

\def\eeeq{\end{eqnarray}} 
 \newcommand{\kvet}{\mbox{\boldmath $k$}}

\def\beeq{\begin{eqnarray}} \def\eeeq{\end{eqnarray}} 
  \def\to{\rightarrow}

  \def\ID{1 \kern -.45 em 1}

\begin{document}\vspace{1.cm}
{\centering {\Large\bf TeV scale Dark Matter and
electroweak radiative corrections}

\vspace{1.cm} { \large{\bf Paolo
Ciafaloni}\footnote{paolo.ciafaloni@le.infn.it} and {\bf Alfredo
Urbano}\footnote{alfredo.urbano@le.infn.it}}\\ {\it INFN - Sezione di Lecce
and Universit\`a del Salento}\\ \centerline{\it Via per Arnesano, I-73100
Lecce, Italy }
\vspace{0.4cm} }
\vspace{0.3cm}
\begin{abstract}
Recent anomalies in cosmic rays data, namely, from the PAMELA Collaboration,
can be interpreted in terms of TeV scale decaying/annihilating dark
matter. We analyze the impact of radiative corrections coming from the
electroweak sector of the standard model on the spectrum of the final
products at the interaction point.
As an example, we consider  virtual one loop
corrections and
real gauge bosons emission in the case of a very heavy vector boson
annihilating into fermions.
We find electroweak
corrections that are  relevant, but not as big as sometimes found in the
literature; we relate this mismatch to the issue of gauge invariance.
At  scales much higher than the symmetry breaking scale, one loop electroweak effects are so big that
eventually higher orders/resummations have to be
considered: we advocate for the inclusion of
these effects in parton shower Monte Carlo models aiming at the description
of TeV scale physics.
\end{abstract}

\section{Introduction}

At the TeV scale and beyond, electroweak (EW) radiative corrections enter into
the realm of nonperturbativity:  one loop corrections relevant for the LHC can
reach the 40 \% level (see for instance \cite{pozzo}). It is
surprising that the same electroweak radiative corrections
%generated by the electroweak sector of the Standard Model
 produce small effects (typically less
than 1 \%)  at LEP that probe the characteristic scale of the theory of 100 GeV
and become huge at energies only 1-order of magnitude bigger. The
reason for this is the presence of energy-growing contributions that, as
has been pinpointed in \cite{bibbia}, are related to the infrared structure
of the theory. More precisely, one loop corrections feature double
logarithmic contributions $\propto \alpha_W\log^2(\sqrt{s}/m_W)$, with $\sqrt{s}$ being the
typical c.m. energy of the process considered, and  $m_W$ the weak scale of
the order of $W$ and $Z$ gauge bosons masses; the weak scale itself acts in this
case as an infrared regulator. Various interesting features of electroweak
radiative corrections at energies much higher than the weak scale have been
studied in the last ten years: noncancellation between real and virtual
contributions, which is a unique feature of weak interactions \cite{BN},
resummation of leading effects \cite{resum}, relevance for phenomenology, and, in
particular, for LHC processes \cite{LHC}.

Recent $e^\pm$ excesses observed by PAMELA \cite{pamela},
FERMI \cite{fermi},  and ATIC \cite{atic} (see also \cite{hess})
can be interpreted in terms of
heavy-mass (1 TeV or more)
dark matter (DM)
annihilation or decay \cite{vari}. Clearly, even if the final products
are initially constituted by, say, an electron/positron pair sharing half
of the c.m. energy
each, radiative virtual corrections and emission of additional
particles in the final state will alter the injection spectrum at the
interaction point.
 Then, the following
 relevant
question arises.

Assuming that physics below the DM mass is the standard
model (SM) one, and assuming that the primary annihilation/decay
process is known, what is the  final products spectrum?

Even if the physics describing the process is assumed to be perfectly
known, the answer to this question is by no means trivial. The usual
approach in the literature is to describe the effect of QCD and QED through
analytical calculations and Monte Carlo generators like \small{PYTHIA}
\cite{pythia}; radiative corrections due to weak gauge bosons are usually
neglected. However, including electroweak effects is important for at least
two reasons. Qualitatively, since all SM particles are charged under the
$\mbox{SU}(2)_{L}\otimes \mbox{U}(1)_{Y}$ group, because of particle radiation the final
spectrum will be composed of {\sl all} possible stable particles, whatever
the primary process. For instance, even if a tree level annihilation  into electron/positron is considered, in the final spectrum also antiprotons will be present. Quantitatively, at the TeV scale and beyond EW corrections of infrared origin are typically as big as the tree level values and cannot therefore
be neglected.
These corrections have been considered in the context of DM signals \cite{dentreal,dentvirtual,serpico}; however corrections growing like $s/m_W^2$ were found, while we find corrections featuring double logarithmic growth. The reasons for these discrepancies are analyzed in Sec. \ref{comp}.

The purpose of this paper is to investigate the impact of EW radiative corrections for possible TeV scale DM signals, namely, trying to contribute to give an answer to the question raised above. Since we focus on the impact of EW corrections, we do not aim at constructing a realistic DM model, nor do we consider the effects of propagation from the interaction point to the detection point. Rather,
we consider a very simple model: a heavy $Z'$ gauge boson corresponding to a   $\mbox{U}'(1)$ group factorized with respect to the SM group. We only consider $Z'$ decay into a fermion-antifermion pair, to which we add a weak gauge boson emission and one loop radiative corrections.
%Moreover, we only consider the radiative corrections effects
% on the primary particle (say, a positron) spectrum.

\section{Gauge boson emission in the soft/collinear region}

We add to the standard model Lagrangian $\mathcal{L}_{SM}$
a vector boson $Z'$  with mass $M$  bigger than $1$ TeV, belonging to  an extra $\mbox{U}'(1)$ gauge symmetry and singlet under the
$\mbox{SU}(3)_{C}\otimes \mbox{SU}(2)_{L}\otimes \mbox{U}(1)_{Y}$ gauge symmetry. The relevant couplings with quarks and leptons are dictated by
the property of gauge invariance. Given the usual  $\mbox{SU}(2)_{L}$ doublet  $L=\left(\nu_{L},e_{L}\right)^{T}$  and the  singlets $e_R,\nu_R$ , we have
\begin{equation}\label{inter}
\mathcal{L}_{int.}=Z'_{\mu}\mathcal{J}_{L}^{\mu}
;\qquad \mathcal{J}_{L}^\mu=f_L
\overline{L}\gamma^\mu L+ f^\nu_R\overline{\nu_{R}}\gamma^\mu\nu_{R} +
f^e_R\overline{e_R}\gamma^\mu e_R,
\end{equation}
with similar expressions  holding for the other families and for quarks. Let us  consider
the case $f^e_R=f^\nu_R=0$, so that the $Z'$ couples to left electron and
neutrino with equal strength $f_L$.

We indicate with $\Gamma_2$ the width for the  process $Z'\to e\Bar{e}$; we wish to calculate the effect of adding one weak gauge boson emission.
As is well known, in the high energy regime $M\gg m_W$  the leading contributions to the three-body width $Z'(P)\rightarrow
e^{+}(p_{1})W^{-}(k)\nu(p_{2})$ are produced by the region of the phase
space where the emitted boson is collinear either to the final fermion or
to the final antifermion; moreover the three-body width is factorized with
respect to the two-body one in this region. Here we show explicitly this factorization and we calculate the expression for the three-body width.

%is factorized with respect to the two body width.
%We are interested in the three bodies decay process $Z'(P)\rightarrow
%e^{+}(p_{1})W^{-}(k)\nu(p_{2})$, where the gauge boson $W^{-}$ is
%collinearly emitted from the positron in the final state.
%In this section we show that
%$\Gamma_2(Z'(P)\rightarrow
%e^{+}(p_{1})\nu(p_{2})$
Let us
show that the phase space factorizes in the region where
the gauge boson momentum $k$ is collinear to the emitting fermion momentum
$p_1$
(the region where $k$ is collinear to $p_2$ can be treated in the same way). In fact
in this region $p_1^2=k^2=0$ implies $p^2=0$ (masses neglected); moreover
we can write
\begin{equation}\label{spaziodellefasi}
\frac{d^{3}\overrightarrow{p}_1}{(2\pi)^{3}2p^{0}_{1}}
\frac{d^{3}\overrightarrow{k}}{(2\pi)^{3}2k^{0}}\delta^{(4)}(P-p_1-p_2-k)
\approx \frac{d^{3}\overrightarrow{p}}{(2\pi)^{3}2p^{0}}
\delta^{(4)}(P-p-p_2)\frac{d^{3}\overrightarrow{k}}{(2\pi)^{3}2k^{0}}\left(\frac{1}{1-x}\right),
\end{equation}
where $P$ is the decaying $Z'$ momentum and
$x$ is the fraction of energy carried away by the gauge boson, $k_0=xP_0$.
The three-body phase space therefore factorizes with respect to the two-body
one in the collinear region\footnote{For convenience,
we include a $1/(2M)$ factor in the definition of $d\Gamma$: $d\Gamma_n=1/(2M)|\mathcal{M}_{n}|^{2} d\Phi_n$, with $d\Phi_n$ being the usual phase space.}:
\be\label{fs} d\Gamma_{3}\approx d\Gamma_2
\frac{d^{3}\overrightarrow{k}}{(2\pi)^{3}2k^{0}}\left(\frac{1}{1-x}\right).
\ee
Furthermore, the amplitude squared factorizes as well, as we will show
now.

\begin{figure}[htb]
\begin{center}
\includegraphics[width=15cm]{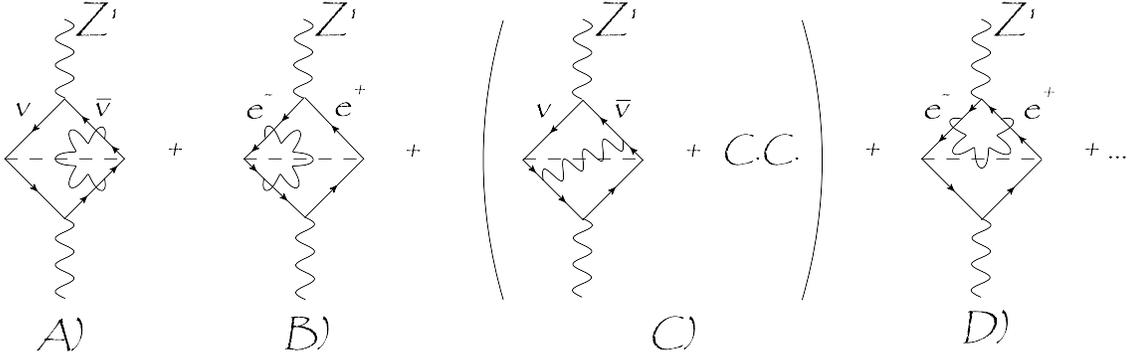}\\
  \caption{Amplitude squared that contributes to the primary particle
  spectrum (see text). (a)-(c) describe real emission contributions,
  while (d) is an example of a virtual correction.}\label{mquadro}
  \end{center}
\end{figure}

Let us first
consider the contribution  to the modulus squared of the
amplitude shown in Fig. \ref{mquadro}C. This contribution can be written as

\begin{equation}\label{C}
|\mathcal{M}|_{C}^{2}=
\frac{f_{L}^{2}g^{2}\epsilon_{\mu}(P)\epsilon_{\rho}^{*}(P)}{[(k+
  p_{1})^2-m_f^2][(k+ p_{2})^2-m_f^2]}D_{\nu\sigma}Tr\left[\slashed{p}_{2}
  \gamma^{\mu}(\slashed{k}+\slashed{p}_{1})\gamma^{\nu}\slashed{p}_{1}\gamma^{\rho}(\slashed{k}+\slashed{p}_{2})\gamma^{\sigma}P_{L}\right],
\end{equation}
where $D_{\nu\sigma}=-g_{\nu\sigma}+k_{\nu}k_{\sigma}/m_{W}^{2}$ is the sum
  over the emitted $W$ physical
  polarizations and $\epsilon_\mu(P)$ is the physical $Z'$ polarization. In the following we will systematically neglect terms
  that, when integrated over the phase space, produce contributions not
  growing with energy: the symbol $\approx$ refers to this
  approximation. Let us first consider the contribution coming from the $g_{\nu\sigma}$
  component of the sum over polarizations. In the collinear
approximation $k\approx x/(1-x) \,p_1$ after some
  Dirac algebra we obtain

\begin{equation}\label{C1}
|\mathcal{M}|_{C_{D_{\mu\nu}=-g_{\mu\nu}}}^{2}
\approx\frac{f_{L}^{2}g^{2}\epsilon_{\mu}(P)\epsilon_{\rho}^{*}(P)}{2(2k\cdot
  p_{1})(2k\cdot p_{2})}\frac{1}{1-x}4p_{1}\cdot p_{2}
Tr\left[\slashed{p}_{2}
  \gamma^{\mu}\slashed{p}_{1}\gamma^{\rho}P_{L}\right],
\end{equation}
which features factorization in the collinear region, since because
of (\ref{fs}):
\be\label{sopra}
%d\Gamma_{C_{\sum_{pol}=-g_{\mu\nu}}}=
 d\Gamma_3|\mathcal{M}|_{C_{-g_{\mu\nu}}}^{2}
=\frac{g^{2}d^{3}\overrightarrow{k}}{(2\pi)^{3}2k^{0}}
\frac{2p_{1}\cdot p_{2}
}{[(k+
  p_{1})^2-m_f^2][(k+ p_{2})^2-m_f^2]}\frac{1}{1-x}
\left\{
\int d\Gamma_2 f_{L}^{2}\epsilon_{\mu}(P)\epsilon_{\rho}^{*}(P)
Tr\left[\slashed{p}_{2}
  \gamma^{\mu}\slashed{p}_{1}\gamma^{\rho}P_{L}\right]\right\},
\ee
and the integral between braces in Eq. (\ref{sopra}) is precisely the tree
level width $\Gamma_2$. In the collinear/infrared region we have (see the
Appendix):
\be
\frac{d^{3}\overrightarrow{k}}{(2\pi)^{3}2\omega}
\frac{2p_{1}\cdot p_{2}
}{[(p_1+k)^2-m_f^2][(p_2+k)^2-m_f^2]}\approx
\frac{dx}{x}\frac{1}{16\pi^{2}}\ln\frac{M^{2}x^2}{4m_W^2},
\ee

so that we finally obtain

\begin{equation}\label{C2integrato}
d\Gamma^C_{\sum_{pol}=-g_{\mu\nu}}\approx
\Gamma_2\quad\frac{\alpha_{W}}{8\pi}\frac{dx}{x}
\left\{4(1-x)\ln\frac{x^{2}M^{2}}{4 m_{W}^{2}}\right\},
\end{equation}
where $\alpha_{W}\equiv g^{2}/4\pi$. Analogous calculations produce the contribution coming from the $k_\mu
k_\nu$ term:
\begin{equation}\label{Cintegrata}
d\Gamma^C_{\sum_{pol}=k_{\mu}k_{\nu}/m_{W}^{2}}\approx
\Gamma_2\frac{\alpha_{W}}{8\pi}\frac{dx}{x}
\left\{2x(x-1)\ln\frac{M^{2}x^{2}}{4m_{W}^{2}}
-\frac{M^{2}x^{2}}{2m_{W}^{2}}\right\}.
\end{equation}
The remaining contributions represented in Figs. \ref{mquadro}(a) and \ref{mquadro}(b) are given by
\begin{equation}\label{A2integrato}
d\Gamma^A_{\sum_{pol}=-g_{\mu\nu}}\approx
\Gamma_2\frac{\alpha_{W}}{8\pi}\frac{dx}{x}
\left\{2x^{2}\ln\frac{M^{2}x^{2}}{4m_{W}^{2}}\right\},
\end{equation}
\begin{equation}\label{Aintegrato}
d\Gamma^A_{\sum_{pol}=k_{\mu}k_{\nu}/m_{W}^{2}}\approx
\Gamma_2\frac{\alpha_{W}}{8\pi}\frac{dx}{x}
\left\{2x(1-x)\ln\frac{M^{2}x^{2}}{4m_{W}^{2}}+\frac{M^{2}x^{2}}{4m_{W}^{2}}\right\},
\end{equation}
and
\begin{equation}\label{B2integrato}
d\Gamma^B_{\sum_{pol}=-g_{\mu\nu}}\approx 0,
\end{equation}
\begin{equation}\label{Bintegrato}
d\Gamma^B_{\sum_{pol}=k_{\mu}k_{\nu}/m_{W}^{2}}\approx
\Gamma_2\frac{\alpha_{W}}{8\pi}\frac{dx}{x}
\left\{\frac{M^{2}x^{2}}{4m_{W}^{2}}\right\}.
\end{equation}
Finally, the sum of all amplitudes squared, generated by Figs. \ref{mquadro}(a)-\ref{mquadro}(c),
 gives

\begin{equation}\label{gauge}
d\Gamma_{\sum_{pol}=-g_{\mu\nu}+k_{\mu}k_{\nu}/m_{W}^{2}}
\approx\Gamma_2P_R^{W}(x)dx;\qquad P_R^{W}(x)=
\frac{\alpha_{W}}{4\pi}\left(\frac{x^{2}-2x+2}{x}\right)\ln\frac{M^{2}x^{2}}{4m_{W}^{2}}.
\end{equation}
Notice that, although terms proportional to $M^2/m_W^2$ are present in the
single contributions (\ref{C2integrato}-\ref{Bintegrato}), these terms
disappear from the final result (\ref{gauge}).

\section{Virtual corrections and primary particle spectrum\label{virt}}
The calculation of virtual corrections, which {\sl must} be included in
order to predict the primary particle spectrum, poses no particular
difficulty.  In the very high energy regime we are considering, they are
dominated by the region of integration over the virtual momentum where the
exchanged gauge boson [see Fig. \ref{mquadro}(d)] is close to the mass shell
and has the same kinematical structure of real emission.  Virtual corrections are
thus dominated by the soft/collinear region just as the real emission ones and
factorized with respect to the tree level amplitude. Referring the reader to
the relevant literature \cite{melles,ewevoeqns}, we wish to point out that,
because of (a) factorization and (b) unitarity of the theory, virtual
contributions can be derived from real emission calculations described in
the previous section.

The fully inclusive decay width $\Gamma_{TOT}(Z'\to f\bar{f}+X)$ is given,
at the order of perturbation theory considered here, by the sum of the
width for the case of no emission [$\Gamma_0(Z'\to f\bar{f})$] and the one
with one gauge boson emitted [$\Gamma_1(Z'\to
f\bar{f}+X,X=\gamma,Z,W)$]. Because of factorization in the leading
collinear/infrared regime, one can write $\Gamma_i=\Gamma_{Born}P_i$, where
$\Gamma_{Born}$ is the tree level value and $P_i$ are functions of
couplings and energy scales. Now, since the theory is unitary, one has
\be\label{V+R}
\Gamma_{TOT}=\Gamma_{Born}(P_0+P_1)=\Gamma_{Born},
\ee so that the inclusive
cross section equals the tree level value and $P_i$ can be interpreted as
probabilities. Now, $P_1$ can be found by integrating $P_R(x)$ in
Eq. (\ref{gauge}) over the
available phase space and
\be\label{VfromR}
P_0=1-P_1=1-\int_0^1 dx P_R^{W}(x)\approx
%1-\frac{\alpha_{W}}{4\pi}\ln\frac{M^{2}(1-z)^{2}}{m_{W}^{2}}\left[\frac{1+z^{2}}{1-z}\right]\vartheta\left(1-\frac{2m_{W}}{M}-z\right)
1-\frac{\alpha_{W}}{4\pi}\left[\frac{1}{2}\ln^{2}\frac{M^{2}}{4m_{W}^{2}}-\frac{3}{2}\ln\frac{M^{2}}{4m_{W}^{2}}\right].
\ee
If we indicate with $z=1-x$ the momentum fraction carried away by the positron,
virtual corrections are described by a distribution peaked at $z=1$, so that  one finally
obtains

\be \label{PV}P_V^{W}(z)=P_0\delta(1-z)
%\left\{1-\frac{\alpha_{W}}{4\pi}\int_{0}^{1-\frac{2m_{W}}{M}}dz'
%\ln\frac{M^{2}(1-z')^{2}}{4m_{W}^{2}}\left[\frac{1+z'^{2}}{1-z'}\right]
%\right\}
=\delta(1-z)\left\{1-\frac{\alpha_{W}}{4\pi}\left[\frac{1}{2}\ln^{2}\frac{M^{2}}{4m_{W}^{2}}-\frac{3}{2}\ln\frac{M^{2}}{4m_{W}^{2}}\right]
\right\}.
\ee
The spectrum of the positron is described by the distribution $P^{W}(z)\equiv
P_R(z)+P_V(z)$. To this one must add the effects coming from radiation of a $Z$ boson, which can be derived in a similar way, and the effects of photon radiation, that we discuss below.

In the case of QED the kinematic is different, since the photon is massless and since collinear singularities are
cut off by the emitting particle mass ($m_{f}$, considered here to be a
fermion). The distribution of emitted photons, derived in the Appendix, is
\be\label{pppp} P_R^{\gamma}(z)=\frac{\alpha}{2\pi}\ln\frac{M^{2}}{4m_{f}^{2}}\left[\frac{1+z^{2}}{1-z}\right]
\vartheta\left(1-z-\frac{2\,\epsilon}{M}\right),
\ee
where $\alpha\equiv e^{2}/4\pi$ and where $\epsilon$ is an infrared regulator, with
dimensions of a mass, having the physical meaning of lowest energy for the
photon. The distribution for virtual corrections can be derived using
(\ref{V+R}) and is
\be P_V^{\gamma}(z)=\delta(1-z)\left\{1-\frac{\alpha}{2\pi}\ln\frac{M^{2}}{4m_{f}^{2}}\int_{0}^{1-\frac{2\epsilon}{M}}dz'\left[\frac{1+z'^{2}}{1-z'}\right]
\right\}=\delta(1-z)\left\{1-\frac{\alpha}{2\pi}\left[2\ln\frac{M}{2\epsilon}-\frac{3}{2}\right]\ln\frac{M^{2}}{4m_{f}^{2}}\right\}.
\ee
The QED distributions depend on the arbitrary parameter $\epsilon$ and are
 divergent in the limit $\epsilon\to 0$, so they are  obviously unphysical. As is well known, the way out is to
introduce a finite resolution $\Delta E$ on the observed hard particle
(say, a positron). The physical meaning is that what is really observed is not a positron alone, but rather a positron together with a soft photon of energy $\omega\le \Delta E$.
We take this into account by substituting $P_V^\gamma+P_R^\gamma$ in the region
$z>1-\frac{2\Delta E}{M}$ with a flat distribution whose integral is the same as
the one of $ P_V^{\gamma}(z)+ P_R^{\gamma}(z)$ in that region:
\be \label{phot}
P^{\gamma}(z)=\frac{\alpha}{2\pi}\ln\frac{M^{2}}{4m_{f}^{2}}\left[\frac{1+z^{2}}{1-z}\right]\vartheta\left(
1-z-\frac{2\Delta E}{M}\right)
+\frac{M}{2\Delta E}\left\{1-\frac{\alpha}{2\pi}\left[2\ln\frac{M}{2\Delta E}-\frac{3}{2}\right]\ln\frac{M^{2}}{4m_{f}^{2}}\right\}
\vartheta\left(z-1+\frac{2\Delta E}{M}\right).
\ee

\begin{figure}[b]
\begin{center}
  \includegraphics[width=12 cm]{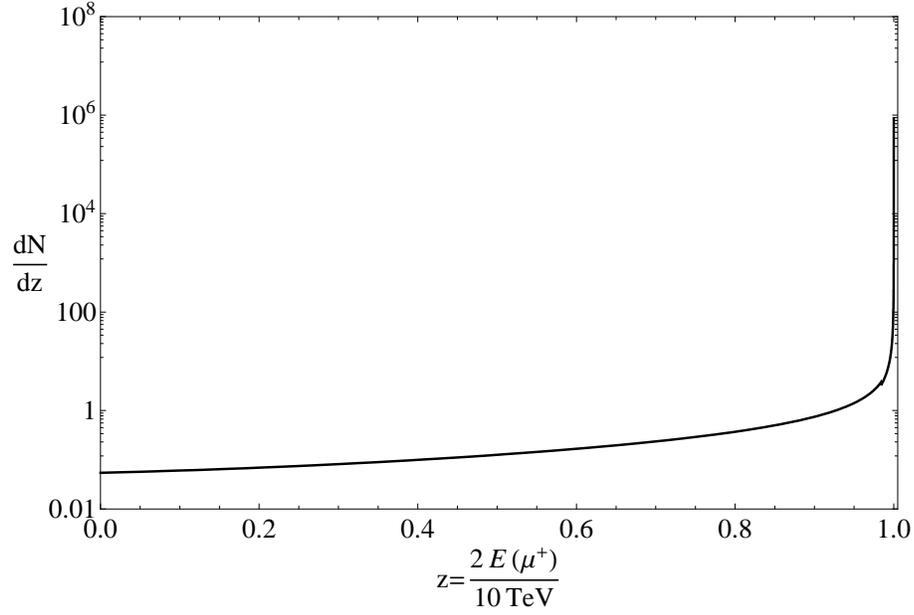}
  \caption{Spectrum $\frac{dN}{dz}$
of the primary particle (antimuon) coming from the
  decay of a heavy $Z'$ ($M=10$ TeV) after inclusion of one loop virtual
  corrections and EW gauge bosons emission. Here and in the following
  figures z$=2E/M$, with $E$ being the antimuon
  energy.
 An ``optimal resolution'' is assumed (see text).}\label{fig:due}
  \end{center}
\end{figure}

\begin{figure}[b]
\begin{center}
  \includegraphics[width=13 cm]{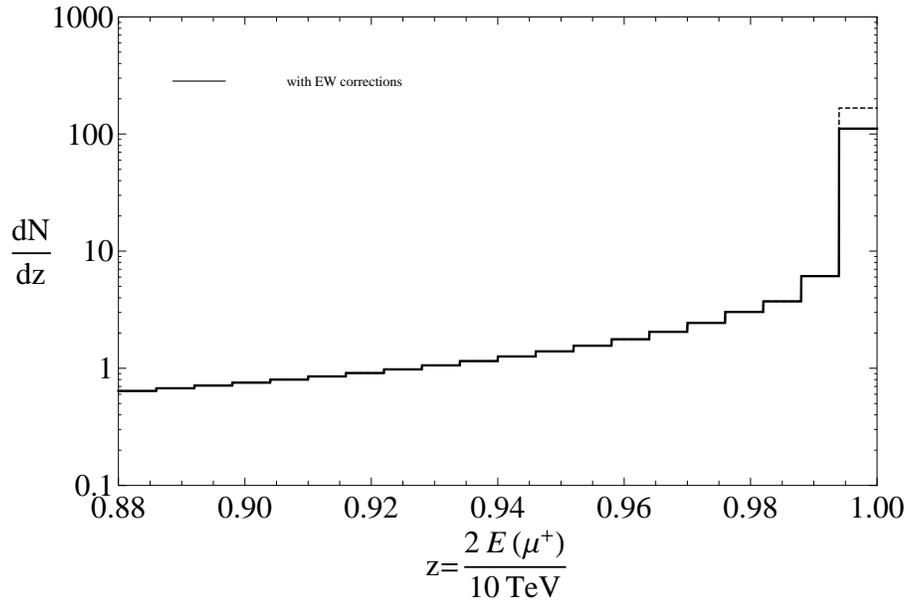}
  \caption{Continuous line: the same as Fig. \ref{fig:due} but with an experimental resolution $\Delta E(\mu)=30$ GeV. Dashed line: tree level distribution with no electroweak corrections.}\label{fig:tre}
  \end{center}
\end{figure}

\begin{figure}[b]
\begin{center}
  \includegraphics[width=12 cm]{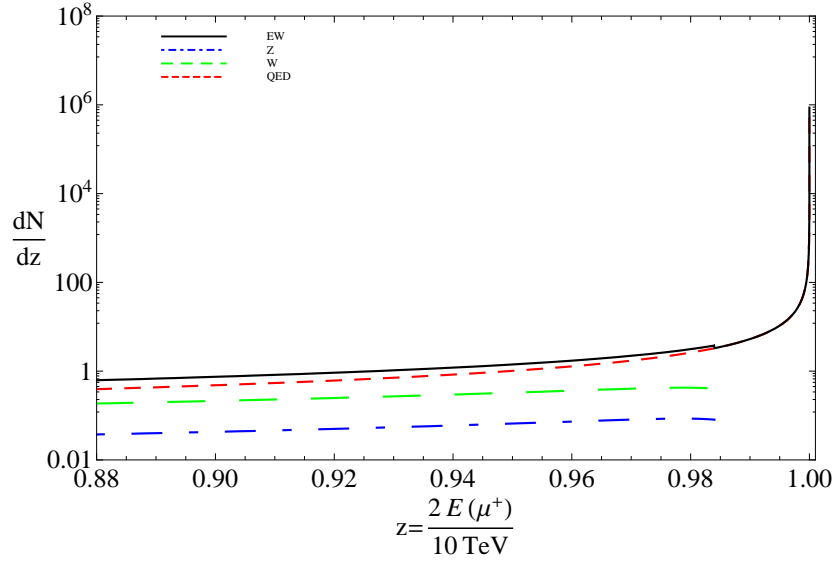}
  \caption{The same as Fig. \ref{fig:due} but zoomed in the region $z\approx 1$ and with the contributions coming from $\gamma$, $W$ and $Z$ emission drawn separately.}\label{fig:zoom}
  \end{center}
\end{figure}

Let us now discuss which value one should give to $\Delta E$. In a collider, this value would be given by the (known) characteristics of the detector. Here however, the situation is less clear because of the effects of propagation from the interaction region to the
point where the detector is physically placed. We choose to analyze two
values for $\Delta E$. The first could be called the ``optimal resolution
case'': $\Delta E$ is chosen in such a way that the distribution
(\ref{phot}) becomes a continuous line; in the case at hand this
corresponds to $\Delta E\approx 0.3$ MeV\footnote{
  Figs. \ref{fig:due},\ref{fig:zoom},\ref{fig:tre} are plotted in the case
  of the spectrum of an
  antimuon and a DM mass $M=10$ TeV.}. The resulting distribution for the positron is drawn in Fig. \ref{fig:due}, where contributions from $\gamma$, $W$ and $Z$ radiation have been added together. Since the actual resolution on the positron energy is certainly worse than the ``optimal'' one, it is possible to obtain the actual distribution from the one in Fig.
\ref{fig:due} simply by dividing it into ``bins'' of finite width. This is
done in Fig. \ref{fig:tre} where the more realistic case $\Delta E$= 30 GeV
has been chosen. Finally, the region $z\approx 1$ from the ``ideal case'' in Fig. \ref{fig:due} is drawn for convenience in
Fig. \ref{fig:zoom}; here the contributions from $\gamma$, $Z$ and $W$ radiation are
drawn separately.

Let us now discuss our results. In the first place, it is apparent from
Fig. \ref{fig:zoom} that the emission of weak gauge bosons plays a
significant role in determining the spectrum of the primary particle;
therefore one should always consider QED ($\gamma$) and weak ($W$,$Z$)
radiation together at very high energies. This is true even more so if the
primary particle is an EW gauge boson itself instead of a fermion. In fact in
this case  QED radiation is partially suppressed because the gauge boson
mass provides the collinear cutoff in Eq. (\ref{pppp}),
in place of the much smaller
fermion mass considered here. Moreover, in the case of final EW gauge
bosons
since EW corrections of infrared
origin are proportional to the Casimir of the external legs representations
\cite{LHC}, they are expected to be bigger in magnitude.

We can see from Fig. \ref{fig:due} that even after inclusion of EW
radiative corrections, the antifermion spectrum is rather sharply
peaked at $z=1$. This can be seen also from Fig. \ref{fig:tre}, where the
tree level distribution (dashed line) is also plotted for
convenience. Virtual EW corrections deplete the first bin (from the right)
by about 30 \%, and produce therefore a significant effect. However the
second bin is depressed with respect to the first one by 1-order of
magnitude,  and the others are even lower, so that the great majority of events
falls into the first bin even after including radiative corrections.

\section{Comparison with existing calculations \label{comp}}

In \cite{dentreal} the effect of adding a weak gauge boson emission to a DM annihilation cross section at very high energies $\sqrt{s}\gg m_W$ was considered; in
 \cite{serpico} a similar calculation for the case of very heavy decaying DM
 ($M\gg m_W$) was done. In both cases, corrections growing like the square of the c.m. energy were obtained. In particular, it was found that
  the cross section (width) with one additional gauge boson in the final state is obtained by multiplying the original one by a factor
 $R_{Z,W} (M)$:

\be\label{boh}
R_{Z,W}(M)=\alpha_WK_{Z,W}\, \frac{M^2}{m_W^2}+...,
%\Gamma_3(P\to p_1+p_2+k)= \Gamma_2(P\to p_1+p_2)\left[\frac{\alpha_{W}}{192\pi}
%\frac{M^{2}}{m_W^{2}}+...\right],
\ee
where $M$ is the relevant high energy scale ($\sqrt{s}$ in the case of scattering, DM mass in the case of decay), is supposed to be much higher than
the weak scale;  $K_{Z,W}$ is a constant that depends on whether a $Z$ or a $W$ is radiated. The dots here stand for terms
that are subleading in the $M\gg m_W$ regime.

On the other hand, when integrated over the final gauge boson phase space (variable $x$),
Eq. (\ref{gauge}) gives

\be\label{totl} \frac{\Gamma_3(Z'\to \bar{e}W^{-}\nu)}{ \Gamma_2(Z'\to e\bar{e})}
=R_W(M)=\frac{\alpha_{W}}{8\pi}\left[
\ln^{2}\frac{M^{2}}{4m_{W}^{2}}-3\ln\frac{M^{2}}{4m_{W}^{2}}+\ldots
\right].
\ee
Similarly, virtual corrections described by (\ref{PV}) grow like the square of the logarithm
of $M^{2}/m_{W}^2$.
Our result therefore disagrees with the results obtained in \cite{dentreal,
dentvirtual,serpico},
where corrections growing like $M^{2}/m_{W}^2$ were found.
Here we wish to point out that the discrepancy is related to
the introduction of dimension-4 operators that break
$\mbox{SU}(2)_{L}\otimes \mbox{U}(1)_{Y}$
gauge invariance, and that one runs into
ambiguous and possibly inconsistent results when trying to calculate
virtual corrections in such framework.
%We believe that this is related to the fact that \cite{dent}
%considers a model in which $SU(2)\otimes U(1)$ gauge invariance is broken
%in the ``soft'' region $M<\omega<\sqrt{s}$ that produces collinear and
%infrared singularities.  However, as we now show, this option should not be
%taken in consideration since it produces wrong results.

\begin{figure}[htb]
\begin{center}
\includegraphics[width=15cm]{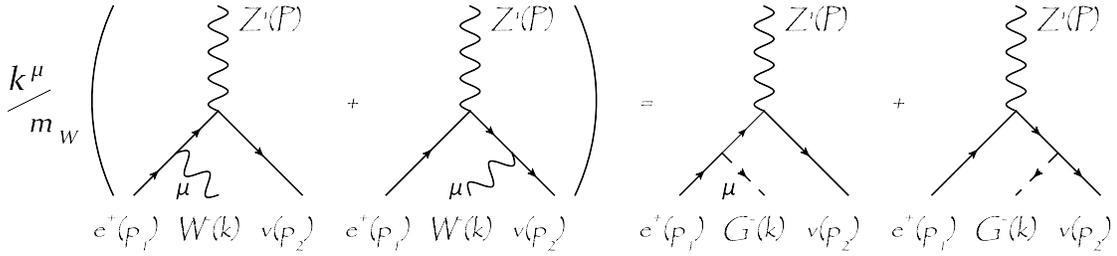}
  \caption{Ward identity relevant for the process considered in this paper.}\label{WIfigura}
  \end{center}
\end{figure}

Terms growing like
$M^{2}/m_{W}^{2}$ are indeed present in the single contributions to the
amplitude squared [see (\ref{Cintegrata},\ref{Aintegrato})] and are related to the terms in $
k_{\mu}k_{\nu}/m_{W}^{2}$ in the sum over the emitted gauge boson
polarizations. However, such terms are absent from the final result
(\ref{totl}). This is a consequence of gauge symmetry in the form of Ward
identities that are depicted in Fig. \ref{WIfigura}. These identities relate
on shell amplitudes with an external gauge boson with corresponding
amplitudes with an external Goldstone, in the following way: \be
\frac{k_\mu}{m_W} {\cal M}_\mu(k,...)=i{\cal M}(\varphi^W(k),...),
 \ee with $m_{W}$
being the mass of the relevant ($W$ or $Z$) gauge boson.\\
 Since Goldstone bosons couple with fermions through their mass, the right-hand side is close to zero.  Therefore the terms in the polarization sum
proportional to $k_{\mu}k_{\nu}$, that are formally dominant by power counting with
respect to the $g_{\mu\nu}$ term, are strongly suppressed at high energies.

A common feature of \cite{dentreal,serpico} is the introduction in the
 Lagrangian of dimension-4 operators that explicitly break $\mbox{SU}(2)_{L}\otimes
 \mbox{U}(1)_{Y}$ gauge invariance: $D_\mu \overline{\nu_L}\gamma^\mu \nu_L$ and/or
 $D(\overline{\nu_R}\nu_L+\overline{\nu_L}\nu_R)$ [to be compared with our gauge
 symmetry invariant Lagrangian in (\ref{inter})].
%Our point of view is that
% this choice undermines the predictivity of the theory and renders the
% calculation of real and virtual corrections unreliable. We relate the lack
% of predictivity in particular to the presence of energy growing
% corrections of infrared origin.
Let us first consider real
 emissions.

Suppose that, following \cite{dentreal}, we introduce a
 $D_\mu \overline{\nu_L}\gamma^\mu \nu_L$ interaction in the Lagrangian without its electron
 counterpart. Clearly, Ward identities are broken: for instance, from the point of view of our gauge invariant example, in the
 diagrams of Fig. \ref{WIfigura}\,\,\,the second one on the left-hand side is
 missing.
 Then, the terms proportional to $M^2/m_W^2$ generated by the
 term
proportional to $k_\mu k_\nu$ in the sum over polarizations  do not cancel
 between the various contributions and are eventually present in the
 expression for the width. What one is really doing here is  largely
overestimating the contribution from longitudinal degrees of
 freedom, whose polarization is $\epsilon_\mu^L=\frac{k_\mu}{m_W}+
O\left(\frac{m_W}{M}\right)$
and whose contribution, although leading by naive power counting with
 respect to transverse, is  suppressed in a gauge invariant theory
because of Ward
 identities.

As we have seen , virtual corrections have to be considered together with
real emission for calculating the observed spectrum and, by virtue of
(\ref{V+R}), are important for unitarity of the theory. However we wish to point out that if, as done in \cite{dentvirtual},
one calculates virtual corrections, the consistency of the calculation
itself is put into doubt because, since  Ward identities are broken,
% in the region
%of energy of the exchanged gauge bosons $m_W<\omega<\sqrt{s}$:
the final results will depend on the (arbitrary) gauge that one chooses in
order to give a prescription for the weak gauge boson propagators. In fact
the infrared structure is
dominated by the region of phase space with on-shell gauge bosons. Since
Ward identities are broken,   the
gauge dependent terms $\propto k_\mu k_{\nu}$ in the propagators are completely
different, say, in the Feynman gauge and in the unitary gauge.
 These problems are generated by the exchange of {\sl soft} quanta with
energies greater than, but close to, the weak scale: therefore, we think
that they
 cannot by cured by
any ``UV completion'' beyond the TeV scale.

%If we insist on a Lagrangian with left neutrino interactions
%without electron counterpart, the situation is confused. In fact, a term
%with electrons is generated by radiative corrections. The presence of such
%a gauge dependent, ill defined on the infrared and UV divergent term
%generated by radiative corrections, does not mean, in our view, that such a
%term can be arbitrarily put to zero, as claimed in \cite{dentvirtual}. On
%the contrary, this means that such a term has to be introduced in the bare
%Lagrangian. Then, as we have argued, the only sensible choice is to choose
%it with a value dictated by gauge symmetry. By not doing so, on top of
%obtaining gauge dependent results for the observables, one is disrupting
%perturbation theory, as is signaled by the presence of radiative
%corrections bigger than tree level values (see
%\cite{dentvirtual,dentreal,serpico}).

\begin{figure}[htb]
\begin{center}
  \includegraphics[width=12 cm]{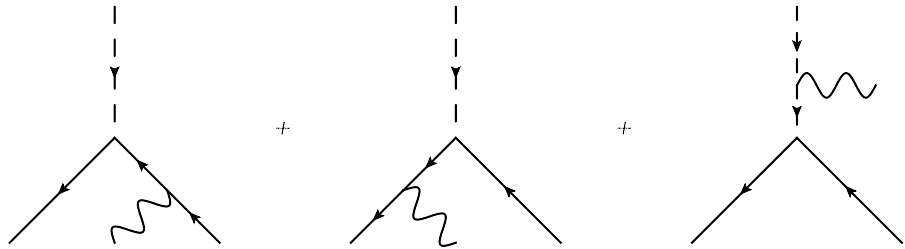}
  \caption{Gauge boson emission diagrams (the wavy line is a $Z$ or a $W$) for the case of a heavy scalar decaying into a fermion or an antifermion.}\label{fig:HD}
  \end{center}
\end{figure}

In Ref. \cite{serpico}, for phenomenological purposes,
a chirality- (and gauge symmetry-)violating
operator of the form $D(\overline{\nu_R}\nu_L+\overline{\nu_L}\nu_R)$, with $D$ being a scalar
and gauge singlet, is introduced into the Lagrangian\footnote{In a gauge
invariant context \cite{serpico2}, a milder growth with energy of
contributions coming from the emission of a gauge boson was found.}. In this case only
tree level gauge boson emissions have been considered.
%If one instead tried
%to calculate radiative virtual corrections, one would possibly run into the kind of ambiguities
%signaled above.
Notice that it is true that radiative
corrections generate chirality violating terms proportional to the gauge
symmetry breaking vacuum expectation value $v$ through fermion masses
\footnote{For some surprising features related to the high energy behavior
of such terms we refer the reader to \cite{amdm}.}. However, since this term
is generated in the framework of a gauge invariant theory (like the SM
itself), Ward identities are obviously respected.  On the contrary, by
introducing a dimension-4 operator that explicitly violates gauge
invariance, Ward identities are broken and by trying to compute radiative
corrections one runs into the problems
signaled above.
One can, of course, write a gauge invariant interaction if the scalar DM particle is an isospin doublet: this would be similar to the usual Higgs-fermion-antifermion coupling in the standard model. We wish to point out that in this case, like in {\sl any} other case where the DM particle has a weak charge, gauge boson emissions from the initial state have to be considered together with emissions from the final legs (see Fig. \ref{fig:HD}). If only final state radiation is considered, Ward identities are broken and one effectively breaks the $\mbox{SU}(2)_{L}\otimes \mbox{U}(1)_{Y}$ symmetry, potentially generating again ``spurious'' terms growing like the square of the energy.

One final comment:
%There is no such a thing as a ``minimal
%branching ratio into electromagnetic final states'':
in our case the $Z'$ can
perfectly well couple to {\sl right} neutrinos only, with the choice
$f_L=f^e_R=0,f^\nu_R\neq 0$.
The $Z'$ then decays only into neutrinos and
this situation, barring possible tiny chirality breaking
effects proportional to fermion masses,
  is left unchanged by radiative corrections. Therefore, no final
  electrically charged states are present, even after including radiative corrections.
%To summarize, if one introduces into the TeV scale
%Lagrangian  hard dimension 4 operators that explicitly
%break $SU(2)\otimes U(1)$ symmetry, one runs into
%inconsistencies, breaking of perturbativity and lack of predictivity of the
%theory.

\section{Conclusions}

In this work we have  analyzed the impact of radiative electroweak  corrections on the spectrum of the final products resulting from the decay of a very heavy ($M>1$ TeV) weakly interacting particle. Determining accurately this spectrum is an important issue, in view of recent experimental results that can be interpreted as a dark matter signal. We have considered a simple model with a $Z'$ gauge boson decaying into leptons, and rediscussed one loop radiative corrections plus emission of a weak gauge boson.
We have found that electroweak corrections play a relevant role in this game; more precisely we can summarize our conclusions as follows:
\begin{itemize}
\item
 EW radiative corrections at one loop
 are of the order of 30 \% in the considered case
 of fermions as primary particles,
 and grow like the log squared of the DM mass. One expects these
 corrections to be even bigger in the case of EW gauge bosons as primary
 particles (see Sec. \ref{virt}); in any case, higher order effects play a
 significant role.
 \item
 QED corrections and ``pure weak'' corrections produced by $W$ and $Z$ exchange have a similar impact on the spectrum of the primary particle, so they should always be considered together.
 \item
 Different from recent results in the literature, we do not find corrections
 growing like the square of the energy. We have shown that the latter
 result can only be obtained if the Ward identities related to
 $\mbox{SU}(2)_{L}\otimes \mbox{U}(1)_{Y}$ gauge symmetry are broken. We have argued that this
 choice can lead to ambiguous results for virtual corrections and we think
 that this point should be better clarified in the future in view of its
 importance.
\item
As already noticed in \cite{serpico}, since EW corrections link all SM
particles, all stable particles (including antiprotons)
will be present in the final spectrum, independently of the primary
particles (say, lepton/antileptons) initially considered.

\item
Heavy DM annihilation/decay can produce final states
 in which no electrically charged particles are present (see Sec. \ref{comp}).
\end{itemize}

Let us now
comment on the possible impact of EW corrections on indirect signals of
dark matter. A detailed quantitative analysis including the effects of
propagation goes beyond the scope of this work, but we can make some
general comments. The first point is related to the fact that the $e^\pm$
excess measured in recent data in the 10 - 100 GeV region
\cite{pamela,fermi,atic,hess}
is not accompanied by a corresponding excess
in the antiproton flux in the same energy region:
 the fraction $\frac{\bar{p}}{p}$ is indeed compatible
with the astrophysical backgrounds \cite{pamela}. One could then consider
for instance DM annihilating to leptons only, or to  $W^+W^-$
pairs with so high an invariant mass that the antiproton excess is shifted to
energies higher than 100 GeV \cite{strumia2}. In both cases, however,
taking into account EW corrections changes the scenario: in the first case
allowing for final antiprotons coming from the decay of the weak gauge
bosons irradiated by the leptons, and in the second case  by softening the
final gauge bosons spectrum. So radiative corrections of EW origin should
be definitely taken in such kind of analyses.

Another  issue is whether it is necessary, for the problem at hand,
to consider higher order EW corrections. At first sight, one loop virtual
corrections plus real gauge boson emission, as are done here, seem to be
sufficient. In fact, as we have seen, one loop corrections are of the order
of 30 \% and higher order virtual corrections should not change the
picture much: the experimental precision of the measurements involved is much
lower than that of a earth based accelerator of course. However, this
conclusion turns out to be a hasty one. In fact we are interested
in  the particles' spectra, in the experimentally interesting region
10-100 GeV, produced by a decay annihilation taking place at a high
invariant mass of the order of 1 TeV or more. Then, even if EW corrections
on, say, the total cross section are relatively small, the impact on the
spectrum at low energies can be dramatic, since ``soft'' ($\sim$ 10 GeV)
particles
 are radiated copiously in the presence of a ``hard ``($\sim$ TeV) process.
 So one can determine whether or not higher order EW corrections are
 relevant in this context only after calculating their effects: this can and
 should be done in the future.
Higher order EW radiative corrections and resummation of leading effects in the high energy regime have been extensively studied in recent years \cite{BN,LHC}; recently
a way of systematizing virtual corrections through soft collinear effective theory has
been considered \cite{scet}. However, EW corrections are usually studied in two
rather ``extreme'' cases: virtual corrections to a given hard process  (in this case, $Z'$ decay into a hard fermion and a hard antifermion) or the inclusive case where soft weak gauge boson emission is fully included. Determining the spectrum resulting from heavy DM annihilation/decay is a different, and difficult, case, where one wishes to know the distribution of a large number of particles in the final state. One possibility to deal with  such a problem is to implement the EW evolution equations \cite{ewevoeqns} (the analogous of the Dokshitzer-Gribov-Lipatov-Altarelli-Parisi equations in QCD) in parton shower Monte Carlo models. This seems to be unavoidable in the nearby future, especially if high energy lepton colliders \cite{lepton}, where showers generated by electroweakly charged particles will be ubiquitous, come closer to see the light.

\section{Appendix: Kinematics}
We consider the decay $Z'(P)\to f(p_1)\bar{f}(p_2)$ with
$P=(M,\mbox{\boldmath{$0$}})$, $p_1=(M/2,\pvet)$, $p_2=(M/2,-\pvet)$, and $p_{1,2}^{2}=m_{f}^{2}$. Then we consider
the additional emission of a gauge boson with momentum $k$ and mass
$\lambda$. The latter is parametrized in the following way:
\be
k=(\omega,\kvet),\quad
k^2=\lambda^2;\qquad\qquad
\pvet\cdot\kvet=|\pvet||\kvet|\cos\theta ;
\ee
 We
consider the region where the emitted boson is collinear to $\pvet_1=\pvet$ so
that $\theta\ll 1$
(the
region where $\kvet$ is collinear to $\pvet_2=-\pvet$ can be treated
in a similar way).
In our calculations the following expression appears:
\be\label{fond}
\frac{d^{3}\overrightarrow{k}}{(2\pi)^{3}2\omega}
\frac{2p_{1}\cdot p_{2}
}{[(p_1+k)^2-m_f^2][(p_2+k)^2-m_f^2]}\,,
\ee
where the differential phase space is evaluated in the collinear region
to give
\be
\frac{d^{3}\overrightarrow{k}}{(2\pi)^{3}2\omega}=
\frac{\omega}{(4\pi)^2}
d\omega d\theta^2.
\ee
The denominator appearing in (\ref{fond}) is evaluated differently in the
case of QED ($\lambda=0$) and in the case of weak gauge bosons emission
($\lambda=m_W\gg m_f$).

In the case of QED we have
\be
(p_2+k)^2-m_{f}^2=2p_2\cdot k\approx 2M\omega,\qquad
(p_1+k)^2-m_f^2=2p_{1}\cdot k\approx
\frac{M\omega}{2}(\theta^2+\frac{4m_{f}^2}{M^2}),
\ee
and the integration over $\theta^{2}$ readily produces

\be\label{fad}
\frac{d^{3}\overrightarrow{k}}{(2\pi)^{3}2\omega}
\frac{2p_{1}\cdot p_{2}
}{[(p_1+k)^2-m_f^2][(p_2+k)^2-m_f^2]}=
\frac{dx}{x}\frac{1}{16\pi^{2}}\ln\frac{M^2}{4m_f^2},
\ee
where $x=2\omega/M$ is the fraction of energy carried away by the emitted
photon.

In the case of weak gauge boson emission we have
\be
(p_2+k)^2-m_{f}^2=m_W^2+2p_2\cdot k\approx 2M\omega,\qquad
(p_1+k)^2-m_f^2=2p_{1}\cdot k+m_{W}^{2}\approx
\frac{M\omega}{2}(\theta^2+\frac{m_{W}^2}{\omega^2}),
\ee
and integrating over $\theta^{2}$:
\be
\frac{d^{3}\overrightarrow{k}}{(2\pi)^{3}2\omega}
\frac{2p_{1}\cdot p_{2}
}{[(p_1+k)^2-m_f^2][(p_2+k)^2-m_f^2]}=
\frac{dx}{x}\frac{1}{16\pi^{2}}\ln\frac{M^{2}x^2}{4m_W^2}.
\ee



\end{document}